\documentclass[12pt]{iopart}

\usepackage[pdftex]{graphicx}
\usepackage{iopams}
\usepackage{epstopdf}
\usepackage{graphicx}
\usepackage{cite}
\usepackage{gensymb}

\begin{document}
\def\be{\begin{equation}}
\def\ee{\end{equation}}
\def\bea{\begin{eqnarray}}
\def\eea{\end{eqnarray}}
\def\rp{r_{+}}
\def\rmm{r_{-}}

\title{
\bf Another Dual Gonihedric 3D Ising Model
}
\date{May 2011}
\author{D. A. Johnston}
\address{Dept. of Mathematics, Heriot-Watt University,
Riccarton,Edinburgh, EH14 4AS, Scotland}

\author{R. P. K. C. M. Ranasinghe}
\address{Department of Mathematics, University of Sri Jayewardenepura,
Gangodawila, Sri Lanka.}


\begin{abstract}

The gonihedric Ising Hamiltonians defined in three and higher dimensions by Savvidy and Wegner
provide an extensive, and little explored, catalogue of spin models on (hyper)cubic lattices
with many interesting features. In three dimensions 
the $\kappa=0$ gonihedric Ising model on a  cubic lattice has been shown to possess 
a  degenerate low-temperature phase and a first order phase transition, as well as interesting dynamical 
properties. The dual Hamiltonian to this may be written as an anisotropic Ashkin-Teller model and
also has a degenerate low-temperature phase as a result of similar symmetries to the original plaquette action.

It is possible to write an alternative dual formulation which utilizes three flavours of spins, rather than the two of the Ashkin-Teller model. This still possesses
anisotropic couplings, but all the interaction terms are now four spin couplings and it
acquires an additional, local gauge symmetry.
We investigate this alternative dual Hamiltonian using zero temperature and mean-field methods together with Monte-Carlo simulations and discuss its properties and the relation to the Ashkin-Teller variant.

\end{abstract} 

\maketitle


\section{Introduction, and One Dual}

The gonihedric Ising models are a family of lattice spin models defined by Savvidy and Wegner 
whose spin cluster boundaries are weighted to mimic
the worldsheets of a gonihedric string action which  was also originally formulated by Savvidy \cite{1}. 
If the string worldsheet is discretized using triangulations, this action may be written as 
\begin{equation}
S = {1 \over 2} \sum_{\langle ij \rangle} | \vec X_i - \vec X_j | \; \theta (\alpha_{ij}),
\label{steiner}
\end{equation}
where
$\theta(\alpha_{ij}) = | \pi - \alpha_{ij} |$, $\alpha_{ij}$ is the dihedral angle between the
neighbouring triangles with a common edge $\langle ij \rangle$
and  $| \vec X_i - \vec X_j |$ are the lengths of the embedded triangle edges.
The word gonihedric is a neologism resulting from the combination of the Greek words gonia
for angle (referring to the dihedral angle) and hedra for base or face (referring
to the adjacent triangles).

The idea of using generalized Ising models to investigate a gas of random surfaces was employed by Karowski \cite{2a} and Huse and Leibler \cite{2b} amongst others
and was described in some detail by Cappi {\it et.al.} in \cite{2}. The plaquettes of the geometrical spin cluster boundaries give the random 
surfaces of interest and one can relate  an Ising Hamiltonian 
with nearest neighbour $\langle ij\rangle $,
next to nearest neighbour $\langle \langle ij\rangle \rangle $ and plaquette $[ijkl]$
terms
\begin{equation}
\beta H = - J_1  \sum_{\langle ij\rangle }\sigma_{i} \sigma_{j} -
 J_2 \sum_{\langle \langle ij\rangle \rangle }\sigma_{i} \sigma_{j} 
- J_3 \sum_{[ijkl]}\sigma_{i} \sigma_{j}\sigma_{k} \sigma_{l}.
\label{e0}
\end{equation}
whose partition function is given by
\begin{equation}
Z = \sum_{\{ \sigma \} } \exp ( - \beta H )	
\end{equation}
to  a surface  partition function for a gas of surfaces
\begin{equation}
Z = \sum_{ \{ S \} } \exp ( - \beta_A A(S) - \beta_I I(S) - \beta_C C (S) )	
\end{equation}
where $S$ is a configuration of spin clusters, $A(S)$ is the total area of boundary plaquettes in spin clusters,
$I(S)$ is the number of edges shared by four plaquettes and $C(S)$ is the number of contiguous plaquettes meeting
and right angles (i.e. ``edges'').  The two sets of couplings are related by
\begin{eqnarray}
\beta_A &=&  2 J_1 + 8 J_2 \nonumber \\
\beta_C &=&  2 J_3 - 2 J_2  \nonumber \\
\beta_I &=&  -4 J_2 - 4 J_3  
\end{eqnarray}
which can be seen by enumerating the possible spin values around the  local plaquette
configurations that give rise to edges, intersections and boundaries.

The gonihedric surface and Ising models are characterized by tuning the area coupling
$\beta_A$ to zero, which fundamentally changes the critical characteristics of the
spin Hamiltonians. This fixes the ratio of nearest neighbour and next to nearest neighbour couplings, so
if we  set the edge coupling to be $\beta$ and ask that
the energy of an intersection, given by convention to be $\beta_I +2 \beta_C$, is $4 \beta \kappa$  we
arrive at the Hamiltonian \cite{3,3a}
\begin{equation}
\label{e1}
H = - 2 \kappa \sum_{\langle ij\rangle }\sigma_{i} \sigma_{j}  +
\frac{\kappa}{2}\sum_{\langle \langle ij\rangle \rangle }\sigma_{i} \sigma_{j} 
- \frac{1-\kappa}{2}\sum_{[ijkl]}\sigma_{i} \sigma_{j}\sigma_{k} \sigma_{l} \; 
\end{equation}
where we have extracted an overall factor of $\beta$ for convenience.
When   $\kappa=0$ the gonihedric Hamiltonian becomes
a purely plaquette term 
\begin{equation}
\label{e2}
H_{\kappa=0} =  - \frac{1}{2} \sum_{[ijkl]}\sigma_{i} \sigma_{j}\sigma_{k} \sigma_{l}
\end{equation}
where the spins live on vertices
and it is the dual(s) of this latter Hamiltonian which are the focus of this paper.

One formulation of the dual to this $\kappa=0$ gonihedric Ising model
was constructed by (Savvidy)$^3$ in \cite{6},
\begin{equation}
\label{dual1}
H = - \frac{1}{2}\sum_{\langle ij \rangle} \sigma_{i}  \sigma_{j} 
- \frac{1}{2} \sum_{\langle ik \rangle } \tau_{i}  \tau_{k} 
- \frac{1}{2} \sum_{ \langle jk \rangle} \eta_{j} \eta_{k} 
\end{equation}
where the sums are carried out over the orthogonal axes $ij, ik$ and $jk$ on the $3d$ cubic lattice
(so each sum is one-dimensional)
and the spin variables on the vertices live in the fourth order Abelian group. For simulations it is perhaps easier 
to deal with Ising ($\pm 1$) spins, which may be done choosing
$\eta_i = \sigma_i \tau_i$ with
$\sigma_i = \tau_i = \pm 1$,  which still satisfies the requisite algebraic relations. The net result is that the
dual Hamiltonian is given by the anisotropic Ashkin-Teller model \cite{7}
\begin{equation} 
\label{dual1b}
H_{Ashkin-Teller} = - \frac{1}{2} \sum_{ \langle ij \rangle} \sigma_{i}  \sigma_{j} 
- \frac{1}{2}  \sum_{ \langle ik \rangle } \tau_{i}  \tau_{k} 
-  \frac{1}{2} \sum_{\langle jk \rangle} \sigma_{j} \sigma_{k} \tau_{j}  \tau_{k} \, .
\end{equation}

In the isotropically 
coupled case the ratio in equ.~(\ref{dual1b}) corresponds to the $\mathbb{Z}_4$ symmetry point
of the standard Ashkin-Teller model
which can be seen by rewriting the Hamiltonian in terms of the four double spins 
$S_i = ( \pm 1, \pm 1)$ to give the 4-state Potts Hamiltonian 
\begin{equation} 
H = - \frac{1}{2} \sum_{ \langle ij \rangle}  \left( 4 \delta_{S_i , S_j }  - 1 \right)
\label{4Potts}
\end{equation}
where the sum now runs over all three edges orientations. This isotropic Hamiltonian has a first order transition
in three dimensions \cite{8}.

We found in \cite{new} that $H_{Ashkin-Teller}$, as for its isotropic Ashkin-Teller counterpart, displayed a 
first order transition. Unlike the isotropic model it had a highly degenerate ground state (and probably low temperature phase) because of the possibility of flipping planes of spins at zero energy cost. Simulations also suggested that its dynamical features  were rather similar to the original
plaquette Hamiltonian, with metastability around the phase transition and, possibly, glassy characteristics
at lower temperatures.

\section{More Duals}

Savvidy and Wegner also considered the formulation of higher dimensional equivalents of the plaquette gonihedric
Hamiltonian and their duals in \cite{6a}.
Borrowing their notation temporarily, a general Hamiltonian in the style of the gonihedric Hamiltonian for $d-n$ dimensional hypersurfaces on a $d$ dimensional hypercubic lattice can be written as
\be
H_{\alpha_1...\alpha_{n+1}}(\vec r) = - K \sum_{\alpha_1<...<\alpha_{n+1}, \vec r} \sum_k
V_{\alpha_1...\alpha_{k-1}\alpha_{k+1}...\alpha_{n+1},
\alpha_k}(\vec r)
\label{ham1}
\ee
where the individual terms are given by the products
\bea
V_{\alpha_1...\alpha_{k-1}\alpha_{k+1}...\alpha_{n+1},
\alpha_k}(\vec r) &=& 
U_{\alpha_1...\alpha_{k-1}\alpha_{k+1}...\alpha_{n+1}}(\vec r)\nonumber \\
&\times&
U_{\alpha_1...\alpha_{k-1}\alpha_{k+1}...\alpha_{n+1}}
(\vec r-\vec e_{\alpha_k}) 
\label{defV}
\eea
with the $U$'s being $-1$ if the dual hyperplaquette belongs to the hypersurface
and $1$ otherwise.
This Hamiltonian is intended to count the number of hyperplaquettes in
the $d-n$ dimensional hypersurface which do not continue straight
through a hyperedge, which is the natural generalization of the three
dimensional gonihedric plaquette Hamiltonian. 

With this choice of Hamiltonian the energy associated with non-straight
hyperedges is $4 K$ times the length of the bend.
To obtain closed surfaces we can again follow the three dimensional case
and introduce Ising spins $\sigma$ attached
to $(d-n+1)$ dimensional hypercubes $\Omega_{\alpha_1...\alpha_{n-1}}$.
The $V$ terms may then be written as the product over the spins at the boundary of two
parallel 
neighbouring $(d-n)$-dimensional hyperplaquettes 
\begin{eqnarray}
V_{\alpha_1...\alpha_{n},\beta}(\vec r) &=& \prod_{k=1}^n
\sigma_{\alpha_1...\alpha_{k-1}\alpha_{k+1}...\alpha_{n}}(\vec r)
\sigma_{\alpha_1...\alpha_{k-1}\alpha_{k+1}...\alpha_{n}}
(\vec r-\vec e_{\alpha_k}) \nonumber \\
&\times&\sigma_{\alpha_1...\alpha_{k-1}\alpha_{k+1}...\alpha_{n}}
(\vec r-\vec e_{\beta})
\sigma_{\alpha_1...\alpha_{k-1}\alpha_{k+1}...\alpha_{n}}
(\vec r-\vec e_{\alpha_k}-\vec e_{\beta}).
\end{eqnarray}
For instance, in four dimensions the resulting Hamiltonian 
for two dimensional hypersurfaces was
\begin{equation}
H^{4d}_{gonihedric}=  - \sum_{P}
(\sigma \sigma \sigma \sigma )_{P}^{~~||}
(\sigma \sigma \sigma \sigma )_{P} \label{4Dham}
\end{equation}
where the sum runs over pairs of parallel
plaquettes $P$ in $3d$ cubes on the $4d$ lattice
and the Ising spins $\sigma$
are now located on the centres of the edges
in the four dimensional lattice.

Several variants of the duals for such Hamiltonians were discussed by Savvidy and Wegner. 
In the case of two dimensional surfaces in general $d$ the dual Hamiltonian was found to be
\be
\label{d2d}
H^{d}_{dual, \, 2d} = -\sum_{\vec r}\sum_{\beta \neq \gamma}
\Lambda_{\beta\gamma}(\vec r)
\Gamma(\vec r,\vec r +\vec e_{\gamma})
\Lambda_{\beta\gamma}(\vec r+\vec e_\gamma).
\ee
where there are $d (d -1 ) /2 $ $\Lambda$ spins at each vertex and  $\Gamma$ spins on each edge.
On the other hand, the dual Hamiltonian for hypersurfaces of codimension one was given by
\bea
H^{d}_{dual, \, codim 1} &=& -K^{*} \sum_{\alpha<\beta, \,  \vec r}\prod_{\gamma}
\Lambda_{\alpha,\beta\gamma}(\vec r)
\Lambda_{\alpha,\beta\gamma}(\vec r+\vec e_{\gamma})
\Lambda_{\beta,\alpha\gamma}(\vec r)
\Lambda_{\beta,\alpha\gamma}(\vec r+\vec e_{\gamma}) \nonumber \\
{} 
\eea
with the standard duality relation relating the coupling to the original $K$
\be
\exp(-2K^{*})=\tanh(2K) \, .
\ee
For two dimensional surfaces in three dimensions either formulation may be employed. 
The codimension one variant gives a dual Hamiltonian with three  different Ising spins $\{\Lambda_{1},
\Lambda_{2},\Lambda_{3} \}$ at every vertex $\vec r$, 
\begin{equation}
\label{L3Ddual}
H^{3d}_{dual, \, codim 1} =  - \sum_{\alpha \neq  \beta \neq \gamma, \, \vec r } \Lambda_{\alpha}(\vec r) \Lambda_{\beta}(\vec r )
\Lambda_{\alpha}(\vec r + \vec e_{\gamma}) \Lambda_{\beta}(\vec r + \vec e_{\gamma}) 
\end{equation}
which at first sight looks rather different to the Ashkin-Teller-like Hamiltonian discussed in the introduction which contains only two spins.
$H^{3d}_{dual, \, codim 1}$ can also be derived from the two-dimensional surface Hamiltonian of equ.~(\ref{d2d}) by summing over the $\Gamma$ edge spins.

Reverting to the notation used in the introduction to facilitate comparisons, let us take
$\Lambda_{1} (\vec r ) = \sigma_i$, $\Lambda_{2} (\vec r ) = \tau_i$ and $\Lambda_{3} (\vec r ) = \mu_i$
which allows us to rewrite equ.~(\ref{L3Ddual}) as
\bea
\label{dual3a}
H_{dual} = - \frac{1}{2}\sum_{\langle ij \rangle} \sigma_i \sigma_j \mu_i \mu_j
- \frac{1}{2} \sum_{\langle ik \rangle } \tau_i \tau_k  \mu_i \mu_k 
- \frac{1}{2} \sum_{ \langle jk \rangle} \sigma_j \sigma_k \tau_j \tau_k 
\eea
where the sums are again carried out over the orthogonal directions $ij, ik$ and $jk$ on the $3d$ cubic lattice and 
for compactness we have denoted this Hamiltonian as simply $H_{dual}$.
Although the Hamiltonian of equ.~(\ref{dual3a}) contains three spins there is a local gauge symmetry
\be
\sigma_i, \, \tau_i, \, \mu_i \, \to  \, -  \sigma_i, \, -\tau_i, \, -\mu_i
\ee
or alternatively
\be
\sigma_i, \, \tau_i, \, \mu_i \, \to \,  \gamma_i  \sigma_i, \,  \gamma_i \tau_i, \, \gamma_i \mu_i
\ee
where $\gamma_i$ is also an Ising spin,
which we would expect to reduce the number of local degrees of freedom back to two. We shall see explicitly how this occurs below.

In the rest of the paper we investigate the behaviour of the dual Hamiltonian $H_{dual}$   
using zero temperature and mean field calculations along with Monte-Carlo simulations and compare it with both the
Ashkin-Teller style dual formulation $H_{Ashkin-Teller}$ of equ.~(\ref{dual1b}) and the original plaquette 
Hamiltonian $H_{\kappa=0}$  of
equ.~(\ref{e2}).

\section{Ground State(s)}

To investigate the ground state (i.e. zero-temperature) structure of the theory
whilst allowing for possible non-uniform states
the Hamiltonian $H_{dual}$ may be written as a sum over the individual cube Hamiltonians $h_C$ \cite{2},
\begin{equation}
h_c =  -\frac{1}{8}\sum_{ \langle ij \rangle} \sigma_i \sigma_k \, \mu_i \mu_j - \frac{1}{8}\sum_{ \langle ik \rangle} 
\tau_i \tau_j   \,  \mu_i \mu_k   - \frac{1}{8} \sum_{ \langle jk \rangle}  \sigma_j \sigma_k \, \tau_j \tau_k \, ,
\end{equation}
where the additional symmetry factor of $1/4$ with respect to the full Hamiltonian is a consequence of one edge being shared by four cubes. Minimizing the energy of the full Hamiltonian can then be achieved by
minimizing the cube energy and tiling the full lattice with compatible minimum energy configurations of individual cubes.

\begin{figure}[h]
\centering
\includegraphics[height=8cm]{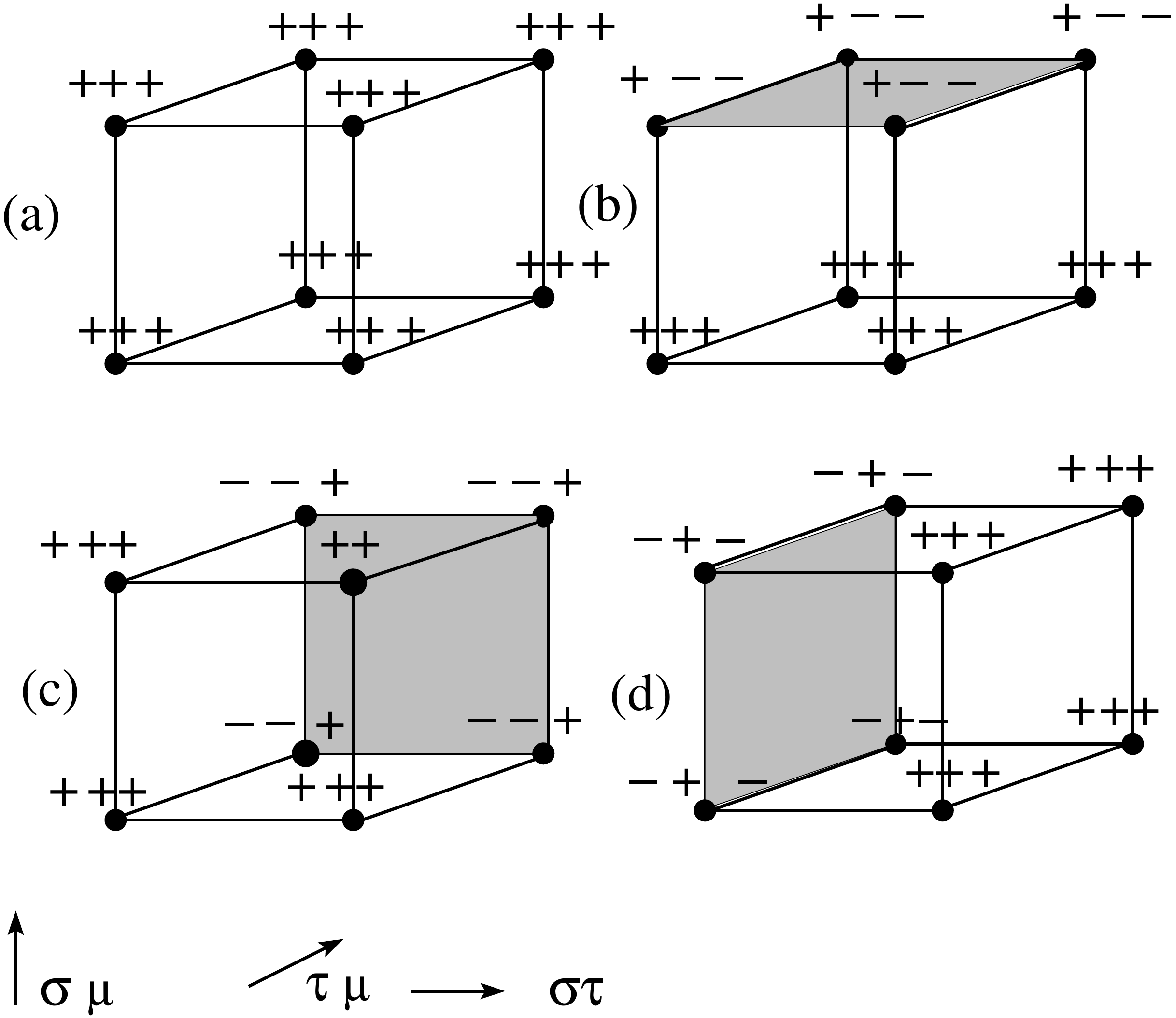}
\caption{Some possible ground state spin configurations on a cube, the $\sigma,\tau,\mu$ values are shown at each vertex. The directions of the anisotropic couplings in the Hamiltonian are indicated, as are the faces on which pairs of spins are flipped. }
\label{groundstate_shade} 
\end{figure}

A feature of both $H_{\kappa=0}$
and its Ashkin-Teller dual $H_{Ashkin-Teller}$
is that it is possible to
flip planes of spins (which may be intersecting) at zero energy cost, giving a peculiar symmetry that lies somewhere
between the gauge and the global. For the dual Hamiltonian under consideration here the anisotropic couplings mean that it is possible to flip planes of {\it pairs} of spins
at zero energy cost as shown in Fig.~(\ref{groundstate_shade}).
It is also possible to flip two or three orthogonal faces on the cube, so tiling the entire lattice with
such combinations we can see that in addition to the purely ferromagnetic ground state we  may have
arbitrary (and possibly intersecting) flipped planes of pairs of spins. In addition, the local gauge symmetry
may also be employed to flip all three spins at any vertex.

\begin{figure}[h]
\centering
\includegraphics[height=8cm]{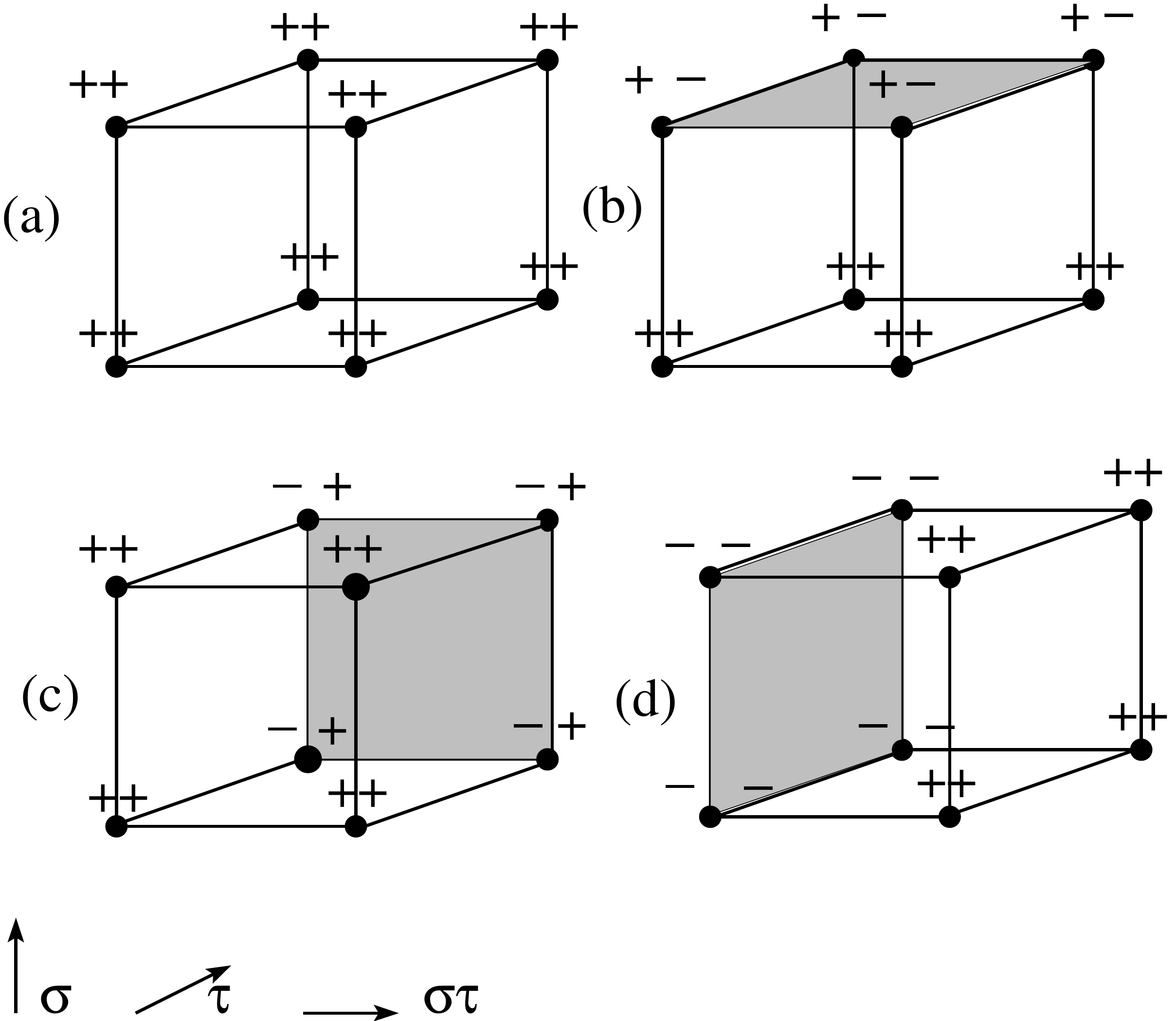}
\caption{Possible ground state spin configurations on a cube for the Ashkin-Teller formulation of the dual
Hamiltonian. The $\sigma,\tau$ values are shown at each vertex. The directions of the anisotropic couplings in the Hamiltonian are again indicated, as are the faces on which spins are flipped. }
\label{groundstateAT_shade} 
\end{figure}

The ground state structure, and the mechanism of anisotropic couplings which allows the plane spin flips,
is clearly very similar to that in $H_{Ashkin-Teller}$ \cite{new}, whose possible ground states we show 
in Fig.~(\ref{groundstateAT_shade}).
We can make the relation even more explicit by using the gauge symmetry to reverse the signs of the spins
on the shaded planes in Fig.~(\ref{groundstate_shade} b,d) which fixes the third spin $\mu$ to be positive
in all configurations. The $\sigma$ and $\tau$ spins then have the same configurations as in 
suitably rotated configurations of the Ashkin-Teller dual.   

In summary, the ground state structure of $H_{dual}$ shows an interesting interplay between the ``semi-local'' planar flip symmetry which appears to be a characteristic of both the gonihedric models and their duals
and a local gauge symmetry. This allows one to reduce the effective number of degrees of freedom and 
recover the ground states of $H_{Ashkin-Teller}$, which not only has anisotropic couplings but a different
form of coupling (energy-energy rather than spin-spin) in one of the directions. All the couplings
in $H_{dual}$, on the other hand, are energy-energy but the price to be paid for this is the 
introduction of the new gauge symmetry. 

\section{Mean Field}

\begin{figure}[h]
\centering
\includegraphics[height=5cm]{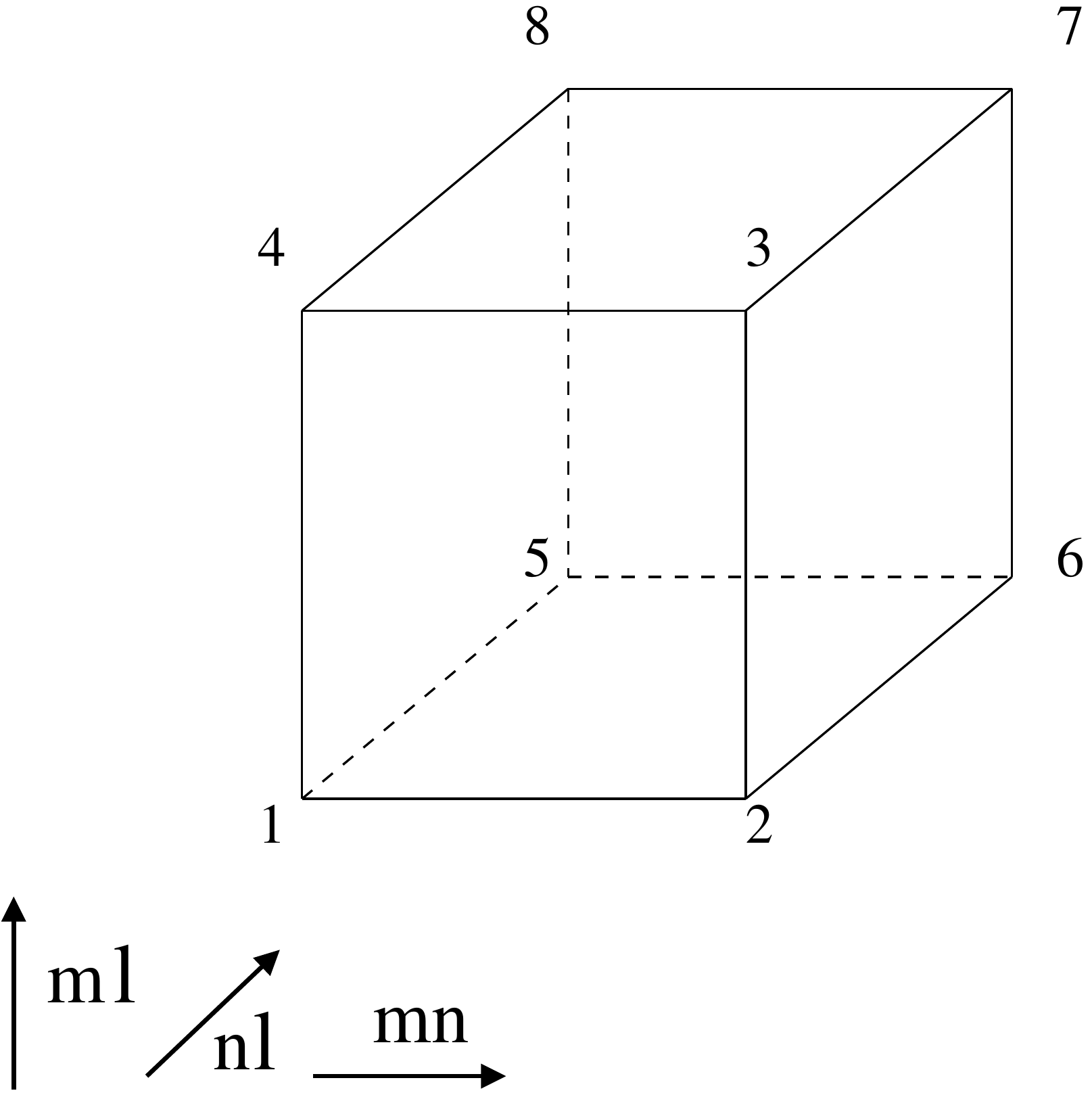}
\caption{The labelling of vertices used in writing the mean field equations for the cube. The directions of the mean field spin couplings in the Hamiltonian are again indicated. }
\label{Fig1} 
\end{figure}

Applying a mean field approach to systems with non-uniform low temperature phases can be done in a similar fashion to the discussion of the ground states by using a cube decomposition \cite{2}.
The total mean field free energy is again written as a sum of the elementary cube terms $\phi(l_{C}, m_{C}, n_C)$, now given by a sum of energy and entropy contributions
\begin{eqnarray}
\beta \, \phi{(l_{C},m_{C},n_C)} &=&  - \frac{\beta}{8}\sum_{ \langle ij \rangle \subset C} m_{i} m_{j} l_{i} l_{k}  -  \frac{\beta}{8}\sum_{ \langle ik \rangle \subset C}  n_{i} n_{k} l_{i} l_{k} \nonumber \\
&-& \frac{\beta}{8}\sum_{\langle jk \rangle \subset C} m_{j} m_{k} n_j n_k \nonumber \\
&+& \frac{1}{16}
\sum_{i \subset C}[(1+l_{i}) \ln(1+l_{i}) + (1- l_{i}) \ln(1 - l_{i})] \nonumber \\
&+& \frac{1}{16}
\sum_{i \subset C}[(1+m_{i}) \ln(1+m_{i}) + (1- m_{i}) \ln(1 - m_{i})] \nonumber \\
&+& \frac{1}{16}
\sum_{i \subset C}[(1+n_{i}) \, \ln(1+n_{i}) \, + \, (1 - n_{i}) \, \ln \,(1 - n_{i})\, ]
\end{eqnarray} 
where the $m_i, n_i, l_i$ are the average  magnetizations corresponding to the spins
$\sigma_i, \tau_i$ and $\mu_i$ respectively and the $\ln$ terms are entropic factors.
We then minimize the cube free energy numerically using the 24 equations
\begin{eqnarray}
\frac{\partial\phi(l_C, m_{C}, n_{C})}{ \partial l_{i}}_{(i=1 {\ldots} 8)} &=& 0  \nonumber \\ 
\frac{\partial\phi(l_C, m_{C}, n_{C})}{ \partial m_{i}}_{(i=1 {\ldots} 8)} &=& 0  \nonumber \\ 
\frac{\partial\phi(l_C, m_{C},n_{C})}{ \partial n_{i}}_{(i=1 {\ldots} 8)} &=& 0
\end{eqnarray}
or, more explicitly
\begin{eqnarray}
m_{1}&=& \tanh[\beta  (m_{4}l_1 l_4  + m_{2} \, n_{1} \, n_{2}) ]\nonumber \\
     &  \vdots& \nonumber \\
m_{8}&=& \tanh[\beta (m_{5} l_5 l_8 + m_{7} \, n_{7} \, n_{8} ) ]  \nonumber \\
     &  \vdots& \nonumber \\
n_{1}&=& \tanh[\beta  (n_{5} l_1 l_5  + n_{2} \, m_{1} \, m_{2}) ]\nonumber \\
     &  \vdots& \nonumber \\
n_{8}&=& \tanh[\beta (n_{4} l_4 l_8  + n_{7} \,m_{7} \, m_{8}) ] \\
 &  \vdots& \nonumber \\
l_{1}&=& \tanh[\beta  (l_{4} m_1 m_4  + l_{5} \, n_{1} \, n_{5}) ]\nonumber \\
     &  \vdots& \nonumber \\
l_{8}&=& \tanh[\beta (l_{4} n_4 n_8 + l_{5} \, m_{5} \, m_{8} ) ] \nonumber  
\label{e2a}
\end{eqnarray}  
where, as is our wont, we have labelled the magnetizations on a face of the cube counter-clockwise $1 \ldots 4$ 
and similarly for the opposing face $5 \ldots 8$,
as shown in Fig.~(\ref{Fig1}).
In disordered phases the mean field magnetizations will iterate to zero, whereas in an ordered
phase they will take non-zero values. 

As in our previous work with such equations stability issues may be addressed by modifying the iterative scheme
slightly from
\begin{eqnarray}
l_i^{(k+1)} = f_i[l^k, m^k, n^k] \, , \; \; 
m_i^{(k+1)} = g_i[l^k, m^k, n^k] \, , \; \;  
n_i^{(k+1)} = h_i[l^k, m^k, n^k] \, , \nonumber \\
{}
\end{eqnarray}
to
\begin{eqnarray}
l_i^{(k+1)} &=&  { \left( f_i[l^k, m^k, n^k] + \alpha l^k_i \right) \over 1 + \alpha} \nonumber \\
m_i^{(k+1)} &=&  { \left( g_i[l^k, m^k, n^k] + \alpha m^k_i \right) \over 1 + \alpha} \nonumber \\
n_i^{(k+1)} &=&  { \left( h_i[l^k, m^k, n^k] + \alpha n^k_i \right) \over 1 + \alpha}
\end{eqnarray}
for a suitable $\alpha$, to deal with the possibility
that it might fail to converge if an eigenvalue of
$ \partial l^{(k+1)}_i / \partial l^k_j \, , \,  \partial m^{(k+1)}_i / \partial m^k_j$
or $ \partial n^{(k+1)}_i / \partial n^k_j$
is less than $-1$ \cite{2}.

The use of coupled equations on a cube accommodates 
non-uniform solutions in an identical manner to the ground state discussion. Any such solution
can then be used to tile the full lattice. If we solve the coupled mean field equations iteratively
a single first-order transition is found in the region of $\beta =0.98$. The ground state chosen depends
on the initial seed values for the $l,m,n$ magnetizations. Choosing these to be near $+ 1$ will lead to the 
simple ferromagnetic ground state, other values will pick out one of the other possible ground states
(or ones related to them by the local gauge transformation that reverses all of the spin signs at a single vertex).
Although the transition temperature calculated from the mean field equations for $H_{dual}$ is different to that observed for the mean field equations in the Ashkin-Teller dual ($\beta=0.83$) we shall see below that Monte
Carlo simulations show there is a much closer similarity between the two systems.

\section{Monte Carlo}

We carry out Monte-Carlo simulations using $10^3, 12^3, 14^3, 16^3, 18^3$ and $20^3$ lattices with periodic boundary conditions for the $\sigma, \tau$ and $\mu$ spins 
 at various temperatures with a simple Metropolis update. After  a suitable number of thermalization sweeps determined by the energy autocorrelation time, $10^7$ measurement sweeps were carried out at each lattice size for each temperature simulated.  

The first observation to make is that there is no signal of a phase transition in the magnetizations
$\langle \sigma \rangle , \langle \tau \rangle$ or $\langle \mu \rangle$, as was also the case with the Ashkin-Teller
dual \cite{new}. Unlike the Ashkin-Teller dual, however, the corresponding susceptibilities for $H_{dual}$ also stay constant,
whereas in the former they showed a sharp drop at the transition point. Given the ability to flip planes of spins
and also reverse the sign of all of the spins at a vertex in a ground state the absence of a simple magnetic order parameter is no surprise, suggesting that these symmetries persist throughout the low temperature phase.

\begin{figure}[h]
\centering
\includegraphics[height=7cm]{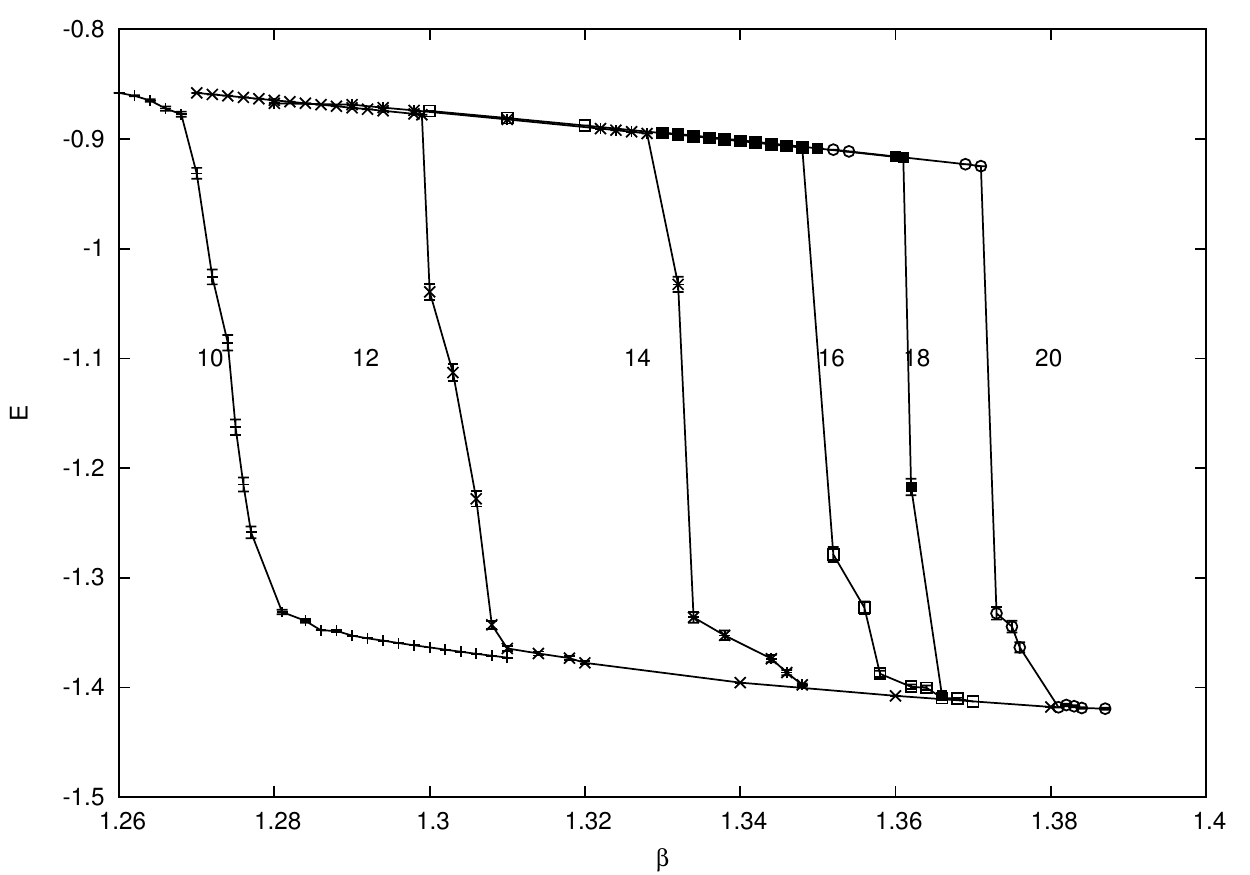}
\caption{The energy for lattices ranging from $10^3$ to $20^3$ from left to right. The lines joining the data points are drawn to guide the eye. Hot starts have been used in the various simulations.}
\label{E0} 
\end{figure}

The phase transition is apparent in measurements of the energy, where there is a sharp drop. A plot of the energy is shown for various lattice sizes simulated in Fig.~(\ref{E0}). 
The energies measured at given inverse temperatures $\beta$ and at given lattice sizes are  
identical to within the error bars for those measured with the Ashkin-Teller Hamiltonian.
We have already seen that the ground states of $H_{dual}$ may be mapped to those of $H_{Ashkin-Teller}$ by a suitable choice of gauge transform, but the two Hamiltonians may be related more generally by carrying out the 
gauge transformations 
\bea
\sigma_i, \, \tau_i, \, \mu_i  \, \to \,   \mu_i \sigma_i, \, \mu_i \tau_i , \, \mu_i^2 
\eea
i.e.
\bea
\sigma_i, \, \tau_i, \, \mu_i \, \to \,  \mu_i \sigma_i, \, \mu_i \tau_i , \, 1  \; .
\eea
This relates the partition functions for $H_{dual}$ and $H_{Ashkin-Teller}$
\bea 
Z &=& \sum_{ \{\sigma,\tau,\mu \}} \exp \left[ - \beta H_{dual} ( \sigma,\tau,\mu) \right] \nonumber \\
&=& 2^{L^3} \sum_{\{ \sigma,\tau \}} \exp \left[ - \beta H_{dual} ( \sigma,\tau,\mu=1) \right] \\
&=&  2^{L^3} \sum_{\{\sigma,\tau \}} \exp \left[ - \beta H_{Ashkin-Teller} ( \sigma,\tau) \right] \; .
\nonumber
\eea
Fixing the gauge in $H_{dual}$ thus corresponds  
to setting $\mu_i =1, \; \forall i$. The result of carrying out this operation,
a unitary gauge fixing,
is to reduce the partition function for the three spin Hamiltonian of $H_{dual}$ to the 
partition function for the two spin Ashkin-Teller dual Hamiltonian.

\begin{figure}[h]
\centering
\includegraphics[height=5cm]{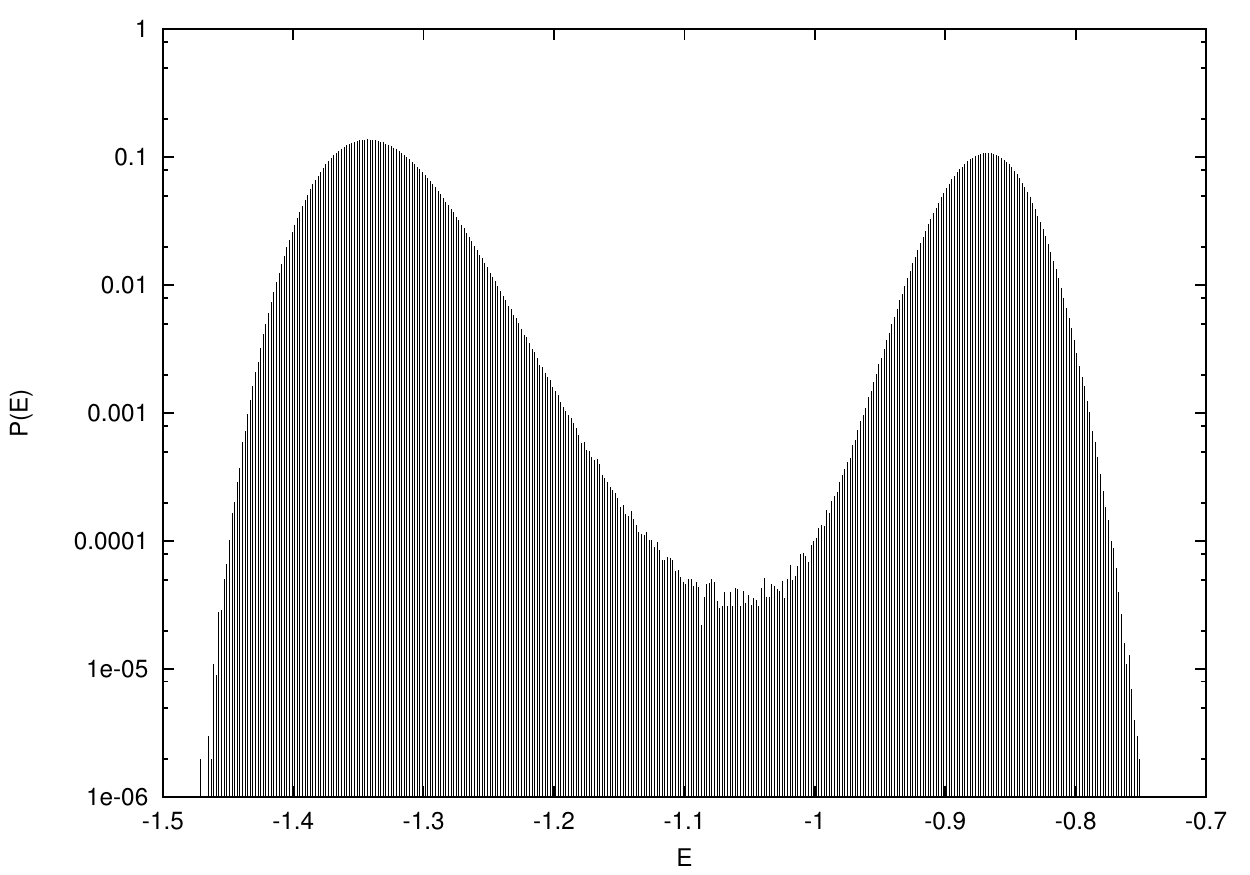}
\caption{The energy histogram from a simulation with $10^7$ sweeps on a $10^3$ lattice near the finite size transition point 
$\beta=1.275$. $P(E)$ is shown on a logarithmic scale.}
\label{bimodal} 
\end{figure}

Since $H_{dual}$  is an ``un-gauge-fixed'' version 
of $H_{Ashkin-Teller}$ we would expect the sharp drop in the energy seen in Fig.~(\ref{E0}) to still be  indicative of a first order transition as it was for $H_{Ashkin-Teller}$. To confirm this we histogrammed the energy during the simulations, which should display a two-peak structure near the finite size pseudo-critical temperature  for  first order transitions.  An example for a $10^3$ lattice close to its finite size pseudo-critical temperature at $\beta_c = 1.275$ is shown in Fig.~(\ref{bimodal}). We have histogrammed the $10^7$ measurements 
of the energy which were carried out after each full lattice sweep of the $\sigma, \tau$ and $\mu$ spins.
The expected double peak structure is clearly visible and is again strikingly similar to that for the Ashkin-Teller dual at the same temperature.

At a first order transition we would also expect to find  a non-trivial limit for Binder's energy cumulant which  is defined as
\begin{equation}
	U_E = 1 - \frac{\langle E^4 \rangle}{3  \langle E^2 \rangle^2} \; .
\end{equation}
This approaches $2/3$ at a second order transition point and a non-trivial limit at a first order point, which is the case here. We can use the finite size scaling of the position of this non-trivial minimum
to obtain an estimate for $\beta_c$ since the expected scaling of the pseudo critical value $\beta_{min}(L)$ is $\beta_{min} (L) = \beta_c - O(1 / L^3)$ at  a first order transition.
If we plot the measured minima positions for the various lattice sizes against $1/L^3$ we get a reasonable fit to this behaviour with a value of $\beta_c \sim 1.391(3)$ and a 
$\chi^2_{dof}$ of $2.25$ when the smallest lattice size of $10^3$ is dropped from the fits.
\begin{figure}[h]
\centering
\includegraphics[height=7cm]{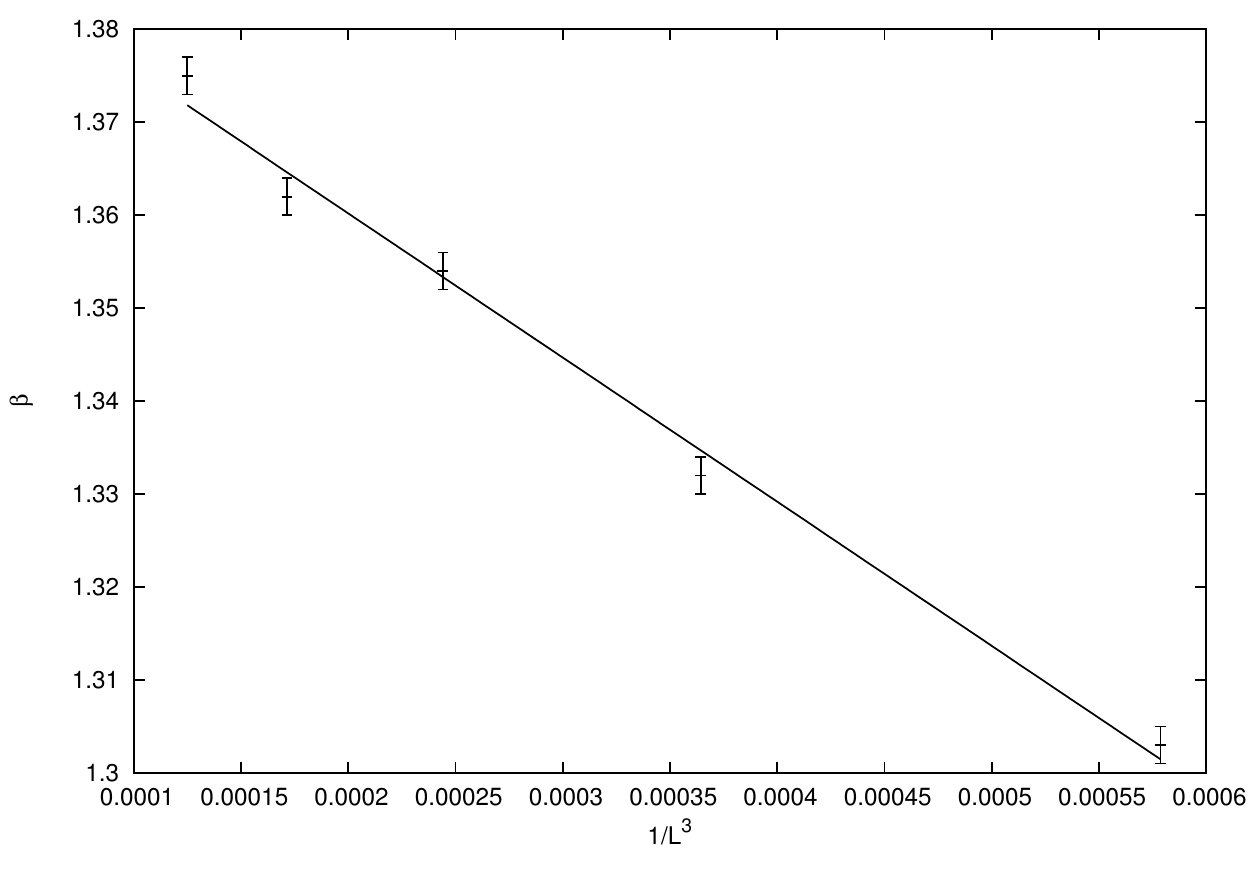}
\caption{The scaling of the position of the minimum of the Binder energy cumulant against the inverse volume.}
\label{E4E2_min} 
\end{figure}
The estimated value of $\beta_c$ from this procedure is consistent with the behaviour of the energy  jump
and an estimate from slow cooling in the next section. It is also consistent with the estimate obtained
using the same procedures for the Ashkin-Teller dual model in \cite{new} of $1.388(4)$. These estimates 
of $\beta_c$ are both rather larger than  an estimate resulting from taking the dual of measurements
of the original plaquette action's critical temperature,  $\beta_c = 0.54757(63)$, in \cite{13}. Allowing for various factors of 2 in the coupling definitions, an estimate for the dual transition temperature is then given by
the standard formula as $ -\ln [ \tanh ( \beta_c/2) ] = 1.32$.  It is possible that the fixed boundary conditions employed in \cite{13}, as well as changing the finite size scaling corrections from the standard periodic boundary conditions,
may have introduced other systematic errors since they have the effect of projecting out the ferromagnetic 
low energy state. The consistency of the estimate for the dual $\beta_c$ from the Ashkin-Teller dual with the estimate here for $H_{dual}$ gives some confidence in the quoted values for the dual(s). Taking $\beta_c=1.391$ gives a dual value
around $0.51$,  lower than the direct estimate in \cite{13} but close to an earlier estimate in \cite{3a}.

\section{Dynamics}

One of the most interesting aspects of the original plaquette Hamiltonian is its dynamical behaviour. It possesses a region of strong metastability around the first order phase transition
and displays glassy characteristics at lower temperatures \cite{5a,5b}. 
We found that the Ashkin-Teller
dual Hamiltonian also appeared to share 
some of these characteristics since it failed to relax to the minimum energy of $E = -1.5$ when cooled quickly from a hot start \cite{new}.
 
\begin{figure}[h]
\centering
\includegraphics[height=7cm]{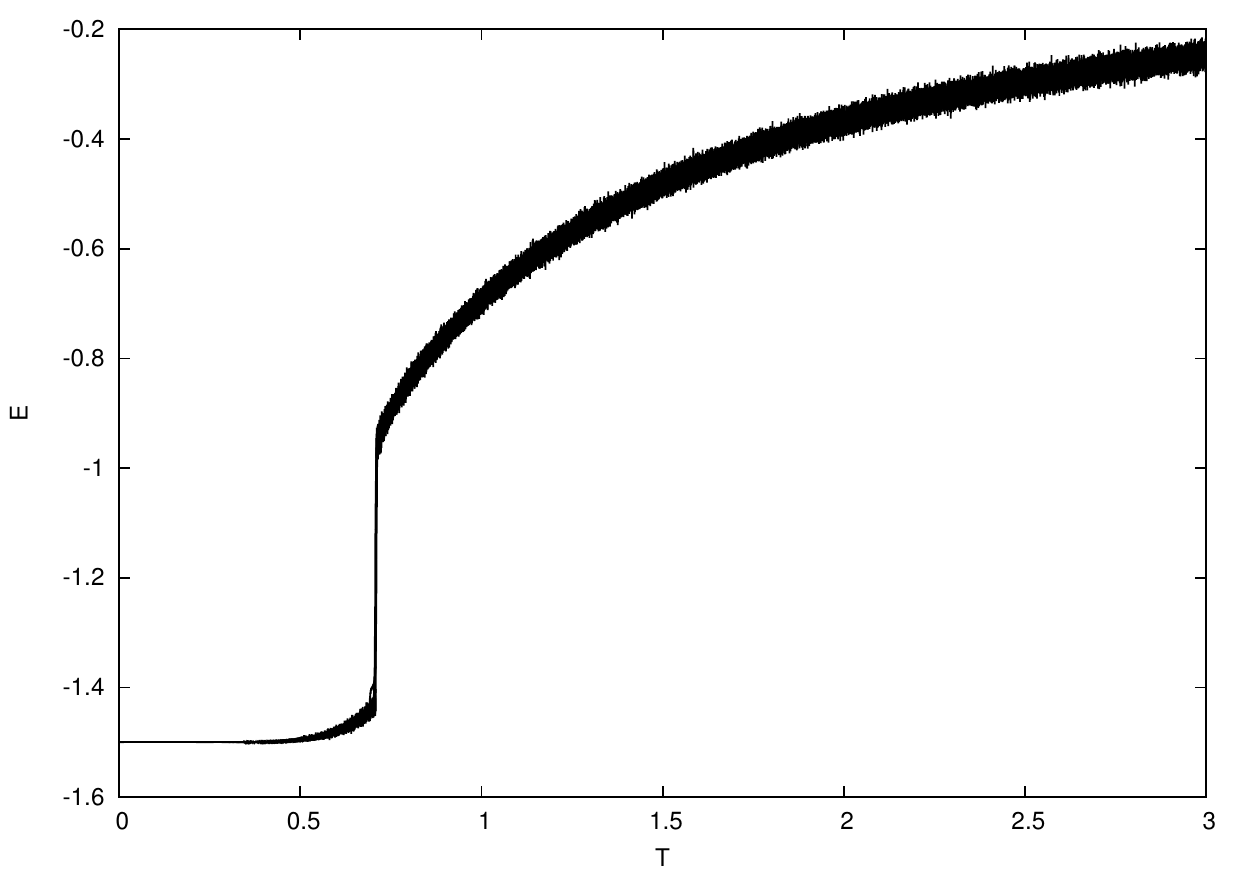}
\caption{The time series of energy measurements obtained from cooling $20^3, 60^3$ and $80^3$ lattices from a hot start at $T=3.0$
at a rate of $\delta T = 0.00001$ per sweep. The traces are effectively indistinguishable.}
\label{r00001} 
\end{figure}

\begin{figure}[h]
\centering
\includegraphics[height=7cm]{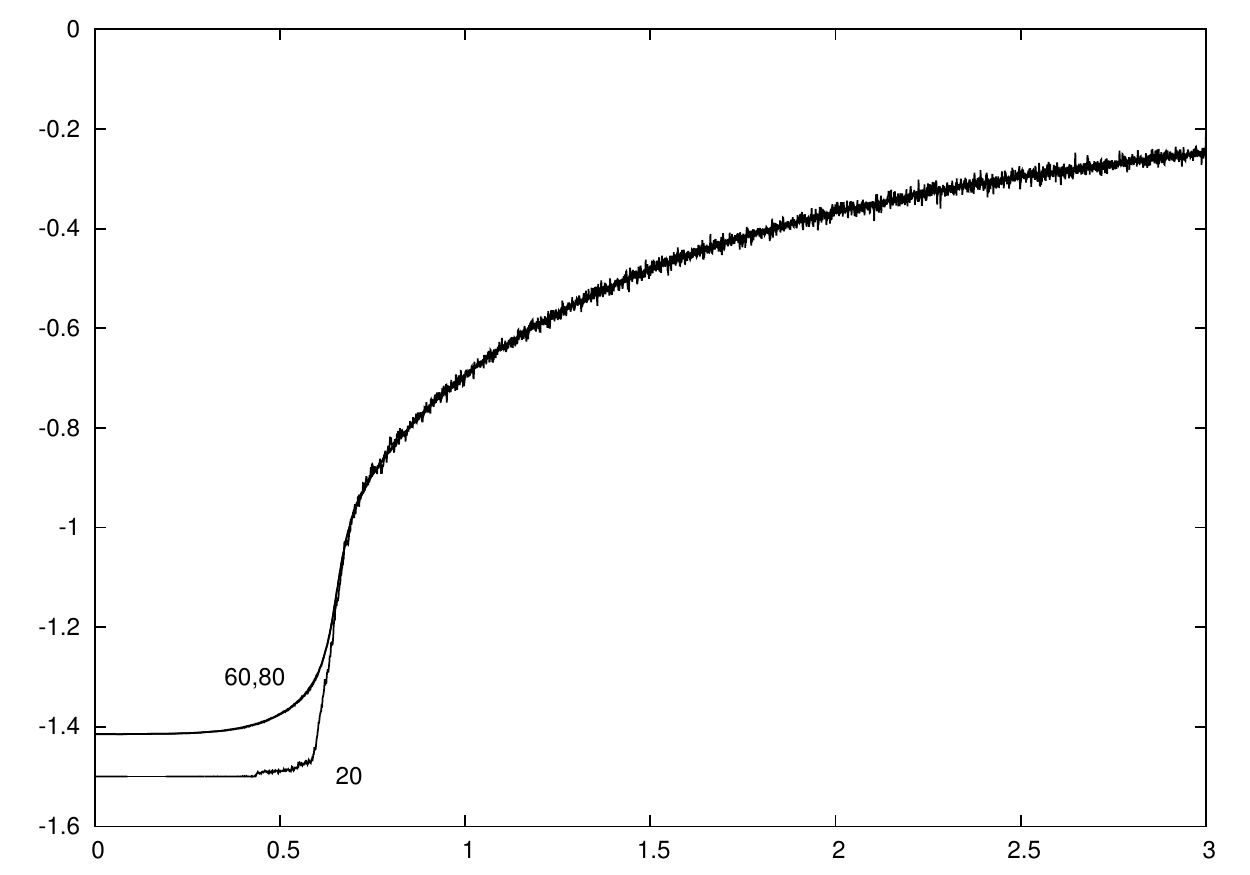}
\caption{The time series of energy measurements obtained from cooling $20^3, 60^3$ and $80^3$ lattices from a hot start
at a rate of $\delta T = 0.001$ per sweep.}
\label{r001} 
\end{figure}

$H_{dual}$ displays identical behaviour under cooling to the Ashkin-Teller dual. We consider
$20^3, 60^3$ and $80^3$ lattices which are first equilibrated at $T=3.0$ and then cooled
at different rates to zero temperature. The energy time series is recorded during this process.
In Fig.~(\ref{r00001}) we can see that with a slow cooling rate of $\delta T = 0.00001$ per sweep, the model still relaxes 
to a ground state with $E=-1.5$ for all the lattice sizes. 
However, as can be seen in Fig.~(\ref{r001}) with a faster cooling rate of $\delta T = 0.001$ per sweep the model no longer relaxes to the ground state energy of $E=-1.5$, but is trapped at a higher value, which is around $-1.415$ for the larger two ($60^3$ and $80^3$) lattices.
The temperature at which the jump in the energy under slow cooling, $T \sim 0.72$ occurs is consistent with the estimate of the transition temperature $\beta \sim 1.391$ arrived at by other means in the previous section.

The plaquette model also presents very similar dynamical behaviour, which has been further investigated in \cite{5a,5b}, showing a region of metastability around the transition point and non-trivial ageing properties
at low temperature.
Whether this is genuinely glassy behaviour or not remains a matter of debate.
It was suggested in  \cite{5b} that in the plaquette model the supercooled high-temperature (``liquid'') phase  becomes physically irrelevant at a temperature below the observed transition point, giving what is effectively an equivalent of the mean field spinodal point.

\section{Discussion}

We have investigated an alternative dual formulation of the plaquette $3d$ gonihedric Ising Hamiltonian of equ.~(\ref{e2}) and clarified its relation with 
the Ashkin-Teller dual of equ.~(\ref{dual1}). An additional gauge symmetry in the Hamiltonian
$H_{dual}$  considered here allows one to reduce the three local degrees of freedom to two
and maps the Hamiltonian onto that of the 
anisotropic Ashkin-Teller dual $H_{Ashkin-Teller}$. There appears to be no simple magnetic order parameter for the low temperature phase in either case because of the flip symmetry and,
for $H_{dual}$, the additional gauge symmetry.
The shared properties of the two dual formulations extend to their dynamics, where fast cooling leads in both cases to a state which has a higher energy than the ground state(s). 

It is a curious feature of both $H_{Ashkin-Teller}$ and $H_{dual}$ here that, although they are dual to an isotropic model, the anisotropic nature of the couplings plays such a fundamental role in determining their properties. We could take ``simplifying'' the Hamiltonian at the expense of introducing
further spins and symmetries a stage further by disentangling
the four spin interactions in $H_{dual}$  by using an additional edge spin $\Gamma^{\alpha}_{ij}, \, \alpha=1,2,3$ acting as an auxiliary field which may be integrated out (or, more correctly summed out) to give 
the codimension one Hamiltonian. This, in effect, amounts to using the two-dimensional surface variant
of the dual Hamiltonian in equ.~(\ref{d2d}) rather than the codimension one variant that gives $H_{dual}$.
This gives a Hamiltonian of the form
\bea
\label{sUs}
H &=& - \frac{1}{2}\sum_{\langle ij \rangle} \left( \sigma_i \Gamma^{(1)}_{ij} \sigma_j  + \mu_i \Gamma^{(1)}_{ij} \mu_j \right) 
- \frac{1}{2} \sum_{\langle ik \rangle } \left( \tau_i \Gamma^{(2)}_{ik} \tau_k +  \mu_i  \Gamma^{(2)}_{ik} \mu_k \right) \nonumber \\
&-& \frac{1}{2} \sum_{ \langle jk \rangle} \left( \sigma_j \Gamma^{(3)}_{jk} \sigma_k  + \tau_j \Gamma^{(3)}_{jk} \tau_k  \right) \; 
\eea 
which we would expect to present identical behaviour to $H_{Ashkin-Teller}$ and $H_{dual}$.
An isotropic version of such a Hamiltonian with one flavour of spin would simply be the gauge-matter coupling
term in the Hamiltonian of the $\mathbb{Z}_2$ gauge-Higgs model which displays no transition on its own, so the anisotropy of equ.~(\ref{sUs}) must again play an important role in determining the phase structure.
 
\section{Acknowledgements}
The work of R. P. K. C. M. Ranasinghe was supported by a Commonwealth Academic Fellowship {\bf LKCF-2010-11}.


\end{document}